\documentclass[9pt,twocolumn,twoside]{forart}

\usepackage{appendix}
\usepackage[english]{babel}

\title{Z-scan applied to phosphate glasses doped with Er$^{3+}-$Yb$^{3+}$ and silver nanoparticles}

\author[1,*]{C. Wiechers}
\author[2]{M. A. Mart\'inez-G\'amez}
\author[3]{M. A. Vallejo-Hern\'andez}
\author[1]{M. Rodr\'iguez -Gonz\'alez}
\author[1]{X. S\'anchez-Lozano}
\author[1]{L. Velazquez-Ibarra }
\author[1]{J. L. Lucio}

\affil[1]{Departamento de F\'isica-DCI, Universidad de Guanajuato, P.O. Box E-143, 37150, Le\'on, Gto., M\'exico.}
\affil[2]{Centro de Investigaciones en \'Optica, Loma del Bosque 115, Col. Lomas del Campestre, Leon 37150, Gto., Mexico}
\affil[3]{Departamento de Ingenier\'ia F\'isica-DCI, Universidad de Guanajuato, P.O. Box E-143, 37150, Le\'on, Gto., M\'exico.}

\affil[*]{Corresponding author: carherwm@fisica.ugto.mx}

\ociscodes{(190.4400) Nonlinear optics, materials; (160.2750) Glass and other amorphous materials;(240.6680) Surface plasmons; (160.5690) Rare-earth-doped materials}

\begin{abstract}

\noindent  We report on the use of Z-scan technique with a beam composed with two modes, to probe the nonlinear optical  pro\-perties of phosphate glasses doped with Er$^{3+}$-Yr$^{3+}$ and silver nanoparticles. Understanding the linear and nonlinear properties of these materials is crucial to evaluate if they are candidates to be used as gain media in lasers or optical amplifiers. Experiments are carried out by implemen\-ting  the open-aperture Z-scan technique with bimodal laser pumping ($LG_{00}$ and $LG_{20}$) at 908.6 nm. The analysis is performed  using a simplified model that incorporates nonlinear absorption and  saturation intensity of the samples. The advantage of using a beam with a bimodal structure is that it allows us to evaluate the energy transfer between the modes, which is relevant since the optical active media act as an intermediary. This process is incorporated through an effective phenomenological parameter in the model that we use in our analysis. \\
\end{abstract}

\begin{document}

\maketitle

\section{Introduction}\label{sec:Int}
Z-scan is a useful experimental technique to characte\-rize optical properties of materials~\cite{Bahae1989, Bahae1990}. Different configurations can be implemented, in particular the open-aperture (OA) and closed-aperture (CA). The former is suitable to study nonlinear absorption and optical saturation, whereas the latter is sensitive to nonlinear refractive index effects. Using high energy pulsed lasers, the Z-scan technique has been implemented to study nonlinear effects far from optical resonances~\cite{Bahae1989}. Pump with continuous wave (CW) lasers at wavelengths around the optical resonances has been reported~\cite{Oliveira1994}. Additio\-nally, it has been used to study ion-doped solids ~\cite{Mendoza1997}. Different theoretical approaches have been proposed to describe the Z-scan results starting from the original description for thin samples and small nonlinear effects~\cite{Bahae1989, Bahae1990}, and many other works which consider large nonlinearities~\cite{Liu2001, Chen2005, Tsigaridas2003, Correa2007, Tsigaridas2009, Ajami2010, Wang2010, Wang2012}, the ellipticity of the beam~\cite{ Mian1996, Tsigaridas2006}, optical sa\-turation~\cite{Bian1999, Fang2000}, and thick samples~\cite{Chapple1994, Tian1995, Liu2003, Palfalvi2009}.\\
Gaussian beams are ideal electromagnetic field configurations; unfortunately, many lasers do not exhibit this characteristic. The Z-scan technique considering higher modes has been addressed in the literature~\cite{Bridges1995, Eriksson1998, Zhang2006}, where they expose that the modes in the Laguerre-Gauss (LG) basis preserve their angular momentum, showing the potential use of the LG basis to deal with non ideal Gaussian beams.  In this work we inspect, from the experimental point of view, the OA Z-scan technique for a beam with two LG modes and we propose a methodo\-logy to analyze the data including the two-mode structure of the pump beam, which improves substantially the fitting of the experimental data, and impacts significantly the extracted va\-lues of the nonlinear properties of the samples. \\
We report an experimental study of nonlinear properties based upon OA Z-scan configuration using CW bimodal pumping at $\lambda_p=908.6$ nm. The advantage of using CW operation is that we can consider that the optical systems in the samples are close to a steady state. The bimodal structure of the beam arises from mode mi\-xing, since we have not attempted to fully filter the fundamental Gaussian mode, which allows us to study the e\-nergy transfer between the spatial modes. \\
We study two phosphate glass matrices: the first contains rare earth (RE) ions ($Yb^{3+}-Er^{3+}$); the composition of the second glass is the same, except that it contains silver nanoparticles (SNP)~\cite{Vallejo2018}.  For both samples the even-order electric susceptibi\-lities are negligible. The RE ions and SNP modify the nonlinear pro\-perties through diverse physical mechanisms, among which we can find: \\ 
$\it{i})$ Yb$^{3+}$ is approximated as a two level system. It is an optical transition, which overlaps with the pump wavelength at $\lambda_p$. Although the Er$^{3+}$ has a complex energy spectrum, it also has some transitions that match with the one in Yb$^{3+}$. Under CW pumping at $\lambda_p$, the excited states $^2 F_{5/2}$ in Yb$^{3+}$ and  $^4I_{11/2}$ in Er$^{3+}$ are  populated affecting the polarizability of the media ~\cite{Powell1990, Velazquez2012}.\\
$\it{ii})$ The Yb$^{3+}$  absorption cross section at $\lambda_p$ is larger than that of Er$^{3+}$, however energy transfer from Yb$^{3+}$ to Er$^{3+}$ increases the Er$^{3+}$ excited state population, enabling two photon absorption in Er$^{3+}$~\cite{Cantelar1998}. \\
$\it{iii})$ The SNP  enhance the electromagnetic field in its neighbourhood. Relevant for our purpose are the broadband of the absorption and emission spectra of the SNP, ensu\-ring the overlap with the dominant $Yb^{3+}-Er^{3+}$ transitions~\cite{Vallejo2018}. Note that third-order nonlinear effects in plasmonic structures have been previously reported~\cite{Husinsky2012, Kindsey2016, Lysenko2016, Lysenko2016b, Garcia2017}. \\
$\it(iv)$ Pumping with a CW laser induces a thermal lens effect, specially when it matches a resonance in active media.\\
$\it(v)$ All dopants may exhibit saturation. \\ 
As far as the physical picture is concerned, the pro\-blem at hand is to describe the light power transmitted through a sample, for different positions of the sample with respect to the field  distribution, for a laser beam whose intensity is position dependent. To this end, we consider a phenomenological model that incorporates nonlinear absorption, optical saturation, and solve for the intensity as a function of the sample position. We firstly apply our approach to the
case of thin samples pumped with a fundamental Gaussian beam and provide an analytic expression for the transmittance. We also discuss a more general si\-tuation, where the pump beam is non-Gaussian and we argue how the thick sample case can be treated. We assume that the dopants in the matrices serve as mediators of energy transfer between the pump modes and we introduce a parameter to characte\-rize this process. The nonlinear parameters that we deal with in this work are interpreted as effective mean values, which are optimized in the model allowing the description of the experimental data.\\
Our goal is twofold: 1) On one hand, it is complementary to the optical sample characterization performed in~\cite{Vallejo2018}. In this regard, our analysis permits the determination of the nonlinear properties of active media. 2) On the other hand, our study implements a methodo\-logy to deal with a bimodal laser  pump. In this context, our results show clear evidence of the relevance of the two-mode coupling, and its impact on the determination of nonlinear properties.\\
This work is structured as follows: in Section~\ref{Sec:Fab} we des\-cribe the samples presenting relevant aspects of their spectra and transitions, as well as their linear optical properties. It also contains the setup and procedure we use to implement the Z-scan technique, and the pump beam composition. The phenomenological approach we use to describe the experiment is presented in Section~\ref{Sec:ZCT}. In Section~\ref{Sec:Res}, we report the expe\-rimental results, the procedure used to fit the data, the discussion of the nonlinear optical characterization of the samples and our conclusions. Finally, in Section~\ref{Sec:Con} we summarize our work. Additionally, we include two appendices regarding technical details of the determination of the li\-near optical properties in Appendix~\ref{App:NA} and the characterization of the beam profile using the LG basis in Appendix~\ref{App:BPC}.

\section{Samples characterization}\label{Sec:Fab}
Two samples of phosphate glass co-doped with $Er^{3+}$ and $Yb^{3+}$ were prepared with equal concentrations; one of them was additionally co-doped with SNP. The SNP are generated as a consequence of  the thermal decomposition of $AgNO_3$ through the reaction: $2AgNO_{3(s)}+\Delta \rightarrow 2Ag_{(s)} + 2NO_{2(g)} + O_{2(g)}$. The sample without SNP, or reference sample, is hereafter referred to as sample A; the sample containing SNP is denoted as sample B. Details regar\-ding the fabrication have been presented in~\cite{Vallejo2018}; here we restrict to the  information relevant to this work. The composition of the samples is specified in Table~\ref{Ta:1}, where the quantities are reported in weight percentage (wt.$\%$).  The glasses were prepared in cylindrical shapes with 7.5\,mm of dia\-meter, and po\-lished upto a thickness $L$ (see Table~\ref{Ta:1}).  

\begin{table}[h!]
\begin{center}
\caption{\bf Composition and thickness of the samples }\label{Ta:1}
\begin{tabular}{lccccc}
 &\multicolumn{4}{c}{\small Concentration (wt. $\%$)} &\small $L$\\
\cline{2-5}
\small Sample &\small  $NaH_2PO_4\cdot H_2O$ &\small  $Yb_2O_3$ &\small  $Er_2O_3$ &\small  $AgNO_3$  &\small   (mm) \\
\hline
\small A &\small  $97.0^{(a)}$ &\small  $2.0^{(a)}$ &\small  $1.0^{(a)}$ &\small  $0.0^{(a)}$ &\small  $1.62^{(b)}$ \\
\small B &\small  $93.0^{(a)}$ &\small  $2.0^{(a)}$ &\small  $1.0^{(a)}$ &\small  $4.0^{(a)}$ &\small  $2.00^{(b)}$ \\
\hline
\end{tabular}
\end{center}
\small $^{(a)}$ The concentration errors are $\pm0.1\%$.\\
\small $^{(b)}$ The thickness errors are $\pm0.01\ \mathrm{mm}$
  
\end{table}

\noindent In Table~\ref{Ta:2} we report measurements of the linear optical pro\-perties. We observe that SNP effects do not affect significantly the linear absorption coefficient, but do change the effective li\-near refractive index. The va\-lues of these optical properties are obtained as described in Appendix~\ref{App:NA}.

\begin{table}[h!]
\centering
\caption{\bf Refractive index and linear absorption coefficient at $\lambda_p=908.6$ nm.}
\begin{tabular}{lcc}
\hline
\small Sample &\small  $n$ &\small  $\alpha$ (mm$^{-1}$) \\
\hline
\small A &\small  $1.649\pm 0.072$ &\small  $0.468\pm 0.026$  \\
\small B &\small  $1.567\pm 0.061$ &\small  $0.458\pm 0.053$  \\
\hline
\end{tabular}
  \label{Ta:2}
\end{table}

\subsection{Spectrum and transitions} \label{Sec:Sam}
\begin{figure}[h!]
\centering
\includegraphics[width=8.5cm]{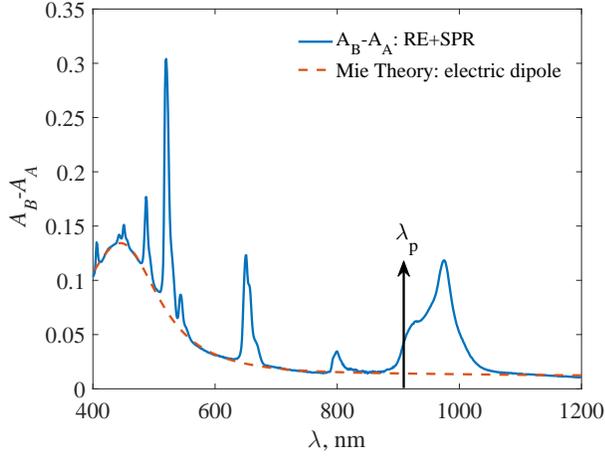}
\caption{\small Difference in absorbance curves between Sample B and Sample A, showing the contributions of the RE transitions and SPR (solid blue line). Theoretical approach using Mie theory with only the electric dipole term (dashed red line). The pump wavelength ($\lambda_p=908.6$ nm) is indicated (vertical black arrow).}
\label{Fi:1}
\end{figure}

The difference in absorbance curves between Sample B and Sample A ($A_B-A_A$) is shown in Fig.~\ref{Fi:1}. The peaks, which are still observed in the $A_B-A_A$ curve, are from $Yb^{3+}$ and $Er^{3+}$ dipolar transitions, since there is an enhancement in the absorption of these transitions due to the SNP (see~\cite{Vallejo2018}). The absorption of SNP is clearly exposed in the $A_B-A_A$ curve. In this way, we ascribe the remnant spectrum to the surface plasmon resonance (SPR), except the peaks. The broadband spectrum of the SPR is expected, since the sizes and shapes of SNP are not uniform. This effect is due to a collection of SPR modes with different wavelength domains. We have used the Mie theory as is described in~\cite{Luk2007}, taking into account the particle size distribution, considering spherical shapes for the SNP and using only their electric dipole contribution. In Fig~\ref{Fi:1}, we also present the SPR spectrum derived from this theoretical approach.  

\begin{figure}[h!]
\centering
\includegraphics[width=8.5cm]{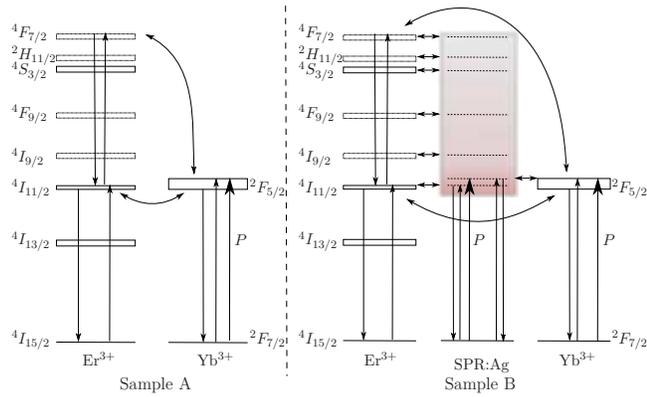} 
\caption{\small Energy level transitions which interact with the pump at 908.6 nm. \textbf{Left:} Sample A dopants exhibit one transition in $Yb^{3+}$ and two sequential level transitions in $Er^{3+}$. The inter-ion energy transfer processes are also depicted. \textbf{Right:} Sample B exhibits the same internal transitions as Sample A. The interspecies energy transfer processes are included. The broadband SPR spectrum allows SNP to interact with the pump and with the RE ions transitions.} \label{Fi:2}
\end{figure}

\noindent A schematic view of the $Er^{3+}$, $Yb^{3+}$ and SPR internal processes is shown in Fig.~\ref{Fi:2}. We sketch the follo\-wing processes related to $\lambda_p$: the following  transitions $^2F_{7/2}\rightarrow\ ^2F_{5/2}$ in $Yb^{3+}$ and  $^4I_{15/2}\rightarrow\ ^4I_{11/2}$ in $Er^{3+}$ , the inter-ion transitions in the $Yb^{3+}- Er^{3+}$ system, and two-photon absorption (TPA) in the $Er^{3+}$. Note that the wide SPR band overlaps with the RE ions absorption peaks, which opens the possibility for the following processes to take place: direct energy transfer among RE ions, and e\-nergy transfer among the RE ions and the SNP. Although we do not attempt to make a detailed microscopic description of the phenomena involved in these systems, we consider it is important to remark that the relevance of these processes rely on their contribution to the nonlinear optical pro\-perties of the samples. 

\subsection{Z-Scan setup and pump beam} \label{Sec:set}
\begin{figure}[h!]
\centering
\includegraphics[scale=0.5]{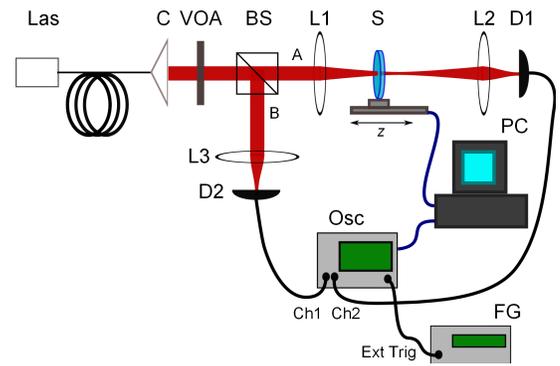}
\caption{\small Z-scan setup: Las- Laser; C - aspherical lens; VOA - variable optical attenuator; BS- Beamsplitter, A - Z-scan path, B - Reference Path; L1 - Lens ($f=3.0\ $cm); L2 and L3 - Lenses ($f=6.0\ $cm); D1 and D2 -Detectors; Osc -Oscilloscope, Ch1 - Channel 1, Ch2 -Channel 2, Ext Trig- External Trigger; FG -Function Generator.} \label{Fi:3}
\end{figure}

\textbf{Z-scan setup.} A schematic representation of our implementation of the OA Z-scan technique is shown in Fig.~\ref{Fi:3}. The source used in the Z-scan system is a CW pigtail diode laser centered at $\lambda_p=908.6$ nm. The fiber output beam is collimated with an aspheric lens (C). The beam power is controlled by a variable optical attenuator (VOA). The beam is divided using a beamsplitter (BS), into beams A and B. The beam A is focused by a lens (L1: $f=3.0\,$~cm) and then collected with a second lens (L2: $f=6.0\,$~cm) into the first detector (D1: Newport 818-BB-20). The space between the lenses L1 and L2 is the region where the Z-scan is performed; the beam A passes perpendicularly through the flat surfaces of the cylindrical samples. The samples are placed in a holder, which is mounted on a displacement stage with a translation  range of  $RZ=3.5\,$~cm,  with steps of $\delta z=200\,\mu$m. The beam B is collected with a third lens (L3: $f=6.0\,$~cm) into the second detector (D2: Newport 818-BB-20). Both detectors outputs are recorded u\-sing an oscilloscope (Osc: Tektronix TDS 3012B). Since the laser ope\-rates in CW mode, the oscilloscope is externally triggered using a Function Gene\-rator (FG: SRS634). 

\noindent The response signals of both detectors ($S_{D1}$ and $S_{D2}$) are characterized beforehand, by verifying their detection stability, linea\-rity dependence on light power, and dark noise. In order to ensure that possible laser fluctuations are corrected, the transmittance is calculated using the ratio between the signals, $T(z)=S_{D1}(z)/(gS_{D2}(z))$, where $g=0.731$ is the correction factor, which ensures $T=1$ if there is no sample loaded in the Z-scan setup. To calculate the Z-scan curves, each $T(z)$ at a fixed z-position is obtained from 5 recorded DC signal traces from the oscilloscope, with 10\,~kilo samples and a full range of 1\,$\mu$s per trace. We then get the mean value and statistical error for each measurement.

\begin{table}[h!]
\begin{center}
\caption{Propagation beam parameters in free-space.} \label{Ta:3}
\begin{tabular}{ccc}
\hline
 \small Mode/m 	&\small  $LG_{00}$/0 &\small $LG_{20}$/2 \\
\hline
\small $w_{0,m}^{2}(\mu \mathrm{m}^{2})$ &\small  $559.4\pm 96.3 $ &\small  $524.3\pm 44.7 $ \\
\small $z_{R,m}(\mathrm{mm})$ &\small  $1.93\pm 0.33 $ &\small  $1.81\pm 0.51 $ \\
\small $f_m(\%)$ &\small  $97.27\pm 0.37$ &\small  $2.73\pm 0.37$\\
\hline
\small $\Delta z_0=z_{0,2}-z_{0,0}(\mathrm{cm})$ &\small $3.09\pm 0.37$ & \\
\hline
\end{tabular}
\end{center}
\end{table}	 
 
\noindent \textbf{Pump beam.} The structure of the beam is taken into account by expanding the intensity beam profile in a Laguerre-Gauss (LG) basis, which is the best suited basis due to the fact that we are working with a pigtail diode laser. The LG modes form a natural basis for cylindrical waveguides. We also remark that LG modes are also suitable to describe Z-scan results, due to their angular momentum conservation. We characterized the pump laser u\-sing the LG basis in free-space finding that its profile is composed by $LG_{00}$ and $LG_{20}$ modes (see Appendix~\ref{App:BPC}). We use these two modes in our experimental Z-scan analysis. The beam propagation parameters and the power fraction of both modes are reported in Table~\ref{Ta:3}. It is worth remarking that the $\Delta z_0$ value measures the distance between the focal planes of the modes.

\section{Transmittance parametrization}\label{Sec:ZCT}
To set our notation and assumptions, we consider the field associated to a CW laser in a single Laguerre-Gaussian or $LG_{ml}$ mode with waist $w_{0}$, traveling through the $+z$ direction,  

\begin{eqnarray}
\small E(r,\phi,z)=&E_{0}C_{ml}\frac{w_{0}}{w(z)}u^{|l|/2}\exp \left(-\frac{u}{2}\right)L_m^{|l|}\left(u\right)\times \nonumber \\
 &\exp \left[ -i\left(\frac{kr^{2}}{2R(z)}+kz+l\phi-\psi(z)\right)\right], \label{Eq:1}
\end{eqnarray}
where $u=2(r/w(z))^2$; $w^{2}(z)=w_{0}^{2}(1+(z-z_{0})^{2}/z^{2}_{R})$ describes the waist evolution in $z$; $R(z)=((z-z_{0})^2+z_{R}^{2})/(z-z_{0})$ is the curvature ratio of the wave front; $z_{R}=kw_{0}^{2}/2$ is the Rayleigh parameter; $z_0$ is the position of the mode focal plane; $k=2\pi n/\lambda_p$ is the magnitude of the wave vector; and $\lambda_p$ is the laser wavelength in vacuum; $n$ is the refractive index;  $E_{0}$ denotes the magnitude of the electrical field at focal plane; $\psi(z)=(2m+|l|+1)\arctan((z-z_0)/z_R)$ is the Gouy phase; and $C_{ml}$ =$\sqrt{2m!/\pi(m+|l|)!}$ is the normalization constant. The field intensity of~\eqref{Eq:1} is,
\begin{equation}
I(r,z)=I(z)|C_{ml}|^2 u^{|l|}\exp \left(-u\right)\left[L_m^{|l|}\left(u\right)\right]^2 \label{Eq:2}
\end{equation}
with $I(z)=I_0(w_{0}/w(z))^2$; $I_0=cn\epsilon_0 |E_0|^2/2$ is the peak intensity at the focal plane; and $c$ is the speed of light in vacuum. The power through any transverse plane is,
\begin{equation}
{\small P_0=\frac{1}{2}\pi I_0w_0^2=\frac{1}{2}\pi I(z)w(z)^2.} \label{Eq:3}
\end{equation}
This relation holds for the peak power and beam width at any transverse plane, in absence of absorbing media.\\
For completeness, we firstly consider the case of a weak field, where the variation of the field intensity across the material is described by the Lambert law of absorption $I_{out}=I_{in}e^{-\alpha L},$ $\alpha$ is the linear absorption coefficient, $L$ is the sample thickness while $I_{in}$ and $I_{out}$ stand for the input and output intensities, respectively. At a given position along the direction of propagation, the power  of the laser beam is obtained by integrating its intensity $I(r,z)$ over the whole transverse plane. The  transmittance is obtained as the ratio between the output and the input po\-	wers at the sample. For a weak field, the linear absorption case, the transmittance reduces to  $T_1=e^{-\alpha L}$; hereafter, we refer to $T_1$ as the linear transmittance. An alternative way to express this result is through  the absorbance $A=-\log_{10}(T)$.

\subsection{Nonlinear electrical susceptibility}
The third-order nonlinear electric susceptibility $\chi^{(3)}$ is a complex quantity, $\chi^{(3)}=\chi_{R}^{(3)}+i\chi_{I}^{(3)}$, which induces a field-intensity dependence in the refractive index and the absorption coefficient:
\begin{eqnarray}
\alpha(I)&=&\alpha+\beta{I},  \hspace{0.5cm}  \beta=\omega\chi_{I}^{(3)}(n^{2}\epsilon_{0}c^{2})^{-1}, \label{Eq:4} \\
n(I)&=&n+\gamma{I},  \hspace{0.5cm}  \gamma=\chi_{R}^{(3)}(2n^{2}\epsilon_{0}c)^{-1}. \label{Eq:5}
\end{eqnarray}
In order to describe the field behavior  within the sample, we determine the position $z'$  dependence of the field intensity and its phase. The OA Z-scan confi\-guration is sensitive to the nonlinear absorption, which gives the field intensity evolution: 
\begin{eqnarray}
\frac{dI}{dz'}&=&-(\alpha+\beta I)I, \label{Eq:6}
\end{eqnarray}

\noindent The $\gamma$ modifies the beam propagation parameters as described by, 
\begin{eqnarray}
\frac{d^2w}{dz'^2}&=&\frac{4}{w^3k^2}\left(1-\frac{\gamma k^2P}{\pi} \right). \label{Eq:7}
\end{eqnarray}
We incorporate the nonlinear refractive index effects through the effective beam propagation parameter ($z_R'$),
\begin{equation}
z_R'=\frac{z_R}{1-\gamma k^2P/\pi}. \label{Eq:8}
\end{equation}
Since $z_R$ evolves into $z_R'$, which is used as a free parameter to fit the data. If the sample is thinner than the effective Rayleigh parameter ($L< z_{R}'$), then we consider the approach of thin samples. There are two options: 1) $\gamma>0 \Rightarrow z_R<z_R'$, 2) $\gamma<0 \Rightarrow z_R>z_R'$.  \\
The solution to~\eqref{Eq:6} is expressed in terms of $I_{in}$ and $I_{out}$, which are the field intensities at the entrance face ($z'=0$) and exit face ($z'=L$), respectively; meanwhile $z$ refers to the sample position:
\begin{equation}
\frac{I_{out}(r,z)}{1+\beta/\alpha I_{out}(r,z)}=\frac{T_1 I_{in}(r,z)}{1+\beta/\alpha I_{in}(r,z)}, \label{Eq:9}
\end{equation}
which we refer to as one of the transverse equations. This relation holds for an arbitrary mode or superposition of modes.

\begin{figure}[h!]
\begin{center}
\includegraphics[width=8.5cm]{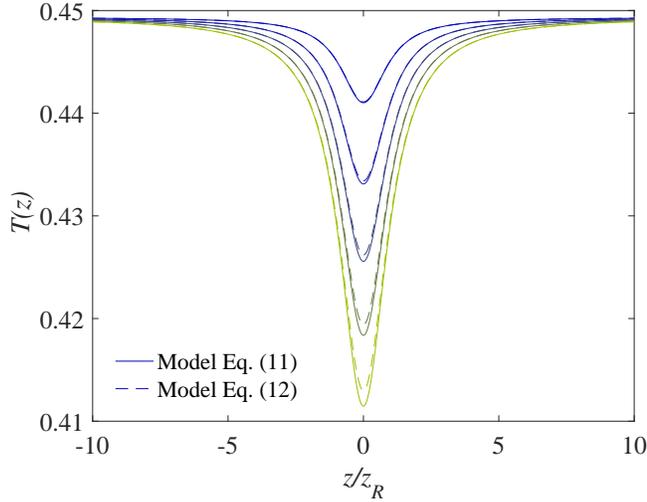}
\caption{\small Nonlinear models: solid lines from~\eqref{Eq:11} and dashed lines from~\eqref{Eq:12}. Parameters values: $L= 2.0\ \mathrm{mm}$, $\alpha = 4.0\ \mathrm{cm}^{-1}$, and $\beta = 100.0 \ \mathrm{cm/MW}$. $P_i$ (mW) = $\{50,\ 100,\ 150,\ 200,\ 250\}$.} \label{Fi:4}
\end{center}
\end{figure}

\noindent The~\eqref{Eq:9} is usually arranged as (see~\cite{Bahae1989, Bahae1990}),
\begin{equation}
I_{out}(r,z)=\frac{T_1I_{in}(r,z)}{1+\beta L_{eff}I_{in}(r,z)}, \label{Eq:10}
\end{equation}	
where $L_{eff}=(1-T_1)/\alpha$. For the $LG_{00}$ mode, the transmittance $T(z)$ is obtained by integrating this equation over the transverse plane at each $z$-position, and divided by the incoming power, $P_{in}$,
\begin{equation}
T(z)=T_1\frac{\ln[1+q(z)]}{q(z)}, \label{Eq:11}
\end{equation}
where $q(z)={\beta}I_{in}(z)L_{eff}$. \\
The following  alternative approach is proposed. Firstly, one can notice that up to the scaling factor $T_1$, both sides of the transverse equation have the same functional form and the relation is fulfilled for each point on the transverse plane. Integrating both sides of~\eqref{Eq:9} over the transverse plane. There is a factor $(w_{out}(z)/w_{in}(z))^2$, which is $\approx 1$ under the thin samples approach. We express the ratio of ${I_{out}(z)}$ to ${I_{in}(z)}$ and obtain the nonlinear transmission:
\begin{equation}
T(z)=\frac{[1+\beta/\alpha I_{in}(z)]^{T_1}-1}{\beta/\alpha I_{in}(z)}. \label{Eq:12}
\end{equation}
Fig.~\ref{Fi:4} shows the comparison among the results obtained of from~\eqref{Eq:11} (solid lines) and~\eqref{Eq:12} (dashed lines) for the transmission curves. The agreement between both results is good, in particular when $\beta /\alpha I<<1$.
\begin{figure}[h!]
	\begin{center}
		\includegraphics[width=8.5cm]{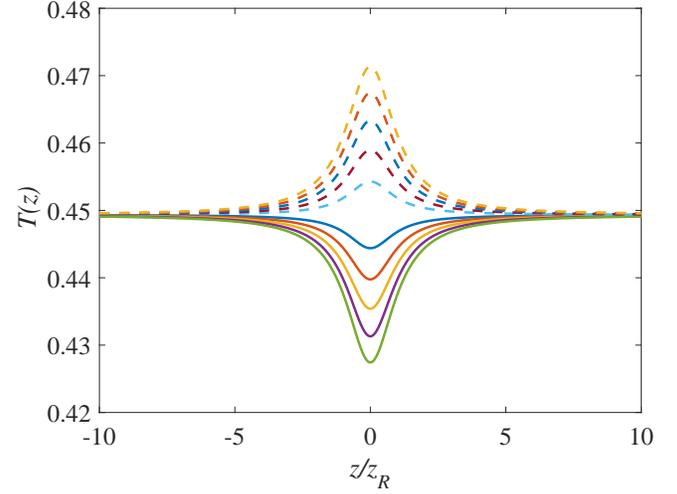}
		\caption{\small Transmittance using the model of~\eqref{Eq:15}. Parameters values: $L= 2.0\ \mathrm{mm}$, $\alpha = 4.0\ \mathrm{cm}^{-1}$, $\beta = 100.0 \ \mathrm{cm/MW}$ and $I_S = 0.1 \ \mathrm{MW/cm}^2$ (solid lines); $\beta = 100.0 \ \mathrm{cm/MW}$ and $I_S = 25.0 \ \mathrm{kW/cm}^2$ (dashed lines). $P_i$ (mW) = $\{50,\ 100,\ 150,\ 200,\ 250\}$.} \label{Fi:5}
	\end{center}
\end{figure}

\subsection{Saturation regime and nonlinear absorption}
For a sample with non-negligible nonlinear properties and sa\-turation intensity, the field evolution within the sample is des\-cribed by:
\begin{equation}
\frac{dI}{dz'}=-\frac{(\alpha+\beta I)I}{1+I/I_S}. \label{Eq:13}
\end{equation}

\noindent Integrating this equation and evaluating at the sample boun\-daries lead to,
\begin{equation}
\frac{I_{out}(r,z)}{(1+\kappa I_{out}(r,z))^{1-\zeta}}=\frac{T_1I_{in}(r,z)}{(1+\kappa I_{in}(r,z))^{1-\zeta}}, \label{Eq:14}
\end{equation}
with $\kappa=\beta/\alpha$ and $\zeta=(\kappa I_S)^{-1}$. For the $LG_{00}$ mode, the alternative procedure we introduce to derive~\eqref{Eq:12} leads to the transmittance,
\begin{equation}
T(z)=\frac{1}{\kappa I_{in}(z)}\left(\left\{1+T_1\left[(1+\kappa I_{in}(z))^\zeta-1 \right] \right\}^{1/\zeta}-1\right). \label{Eq:15}
\end{equation}
The transmittance exhibits either hills or valleys, depending on the values of the parameters ($\beta,\ I_S$) or ($\kappa,\ \zeta$). Some examples are shown in Figure~\ref{Fi:5}. Moreover, in the asymptotic limit $I_S \rightarrow \infty$, the~\eqref{Eq:12} is recovered. 

\begin{figure}[h!]
	\begin{center}
		\includegraphics[width=8.5cm]{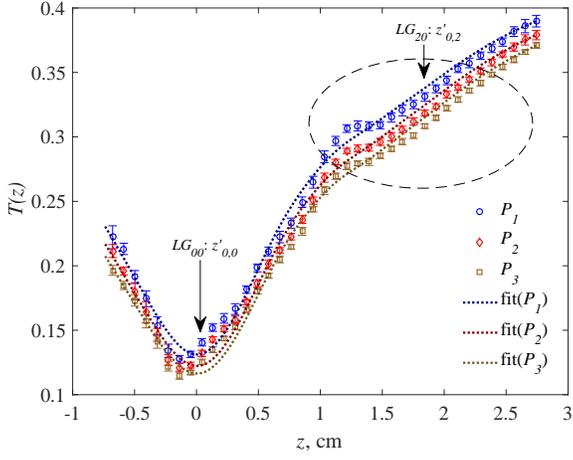}
		\caption{\small Z-scan results Sample A: Measurements (symbols) and fitting curves (dotted lines). Incoming beam pump powers: $P_{in,j}$ (mW) =$\{(244.4\pm 0.7),\ (291.9\pm 0.9),\ (330.3\pm 1.0)\}$. The focal plane positions of the modes are indicated.} \label{Fi:6}
	\end{center}
\end{figure}

\section{Results and Discussion}\label{Sec:Res}
\begin{figure}[h!]
	\begin{center}
		\includegraphics[width=8.5cm]{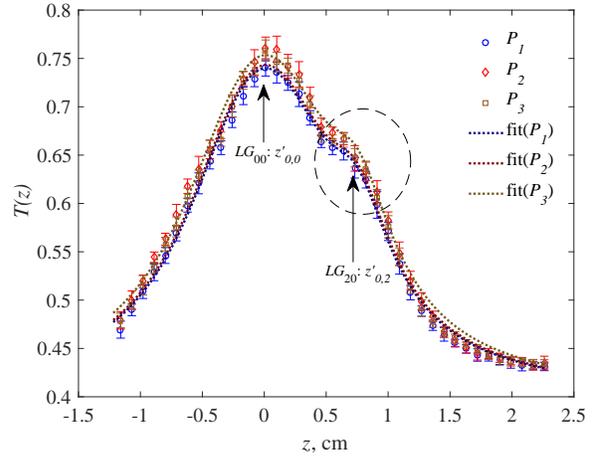}
		\caption{\small Z-scan results Sample B: Measurements (symbols) and fitting curves (dotted lines). Incoming beam pump powers: $P_{in,j}$ (mW) =$\{(262.1\pm 0.9),\ (270.2\pm 0.8),\ (297.3\pm 0.9)\}$. The focal plane positions of the modes are indicated.} \label{Fi:7}
	\end{center}
\end{figure}

Before discussing the fit and the interpretation of the results, further specifications are required. The beam is composed by two Laguerre-Gauss modes, which co-propagate through the samples. We consider that both modes interact independently with the active medium, producing an effective energy transfer between them. This phenomenon is incorporated through the parameter $X$, which is the effective fraction of power transfer between the modes. Thus, the laser intensity is described as the sum of two concentric interacting modes ($LG_{00}$ and $LG_{20}$):
\begin{equation}
\small I_{v,j}(r,z)=\frac{2nP_{v,j}}{\pi}\sum_{m=0}^{1}\frac{(f_{2m}+(-1)^{m+1} X)}{ w_{2m}^2(z)}e^{-u_m}\left[L_{2m}\left(u_m\right)\right]^2. \label{Eq:16} 
\end{equation}
where $u_m=2(r/w_{2m}(z))^2$, $w_{2m}^{2}(z)=\lambda_p(n\pi z'_{R,2m})^{-1}\ [z'^{2}_{R,2m}+(z-z'_{0,2m})^2] $, and $v=\{in,out\}$ since the input and output fields have the same profile but different power. The transmittance is then calculated from: 
\begin{equation}
T_j(z)=\frac{P_{out,j}(z)}{P_{in,j}}. \label{Eq:17}
\end{equation}
where $z$ is the sample position and $P_{in,j}$ remains invariant under changes in $z$.\\
We perform the fitting of the experimental data and the theore\-tical transmittance~\eqref{Eq:17} by means of a least squares method (LSM), where we use the transverse equation shown in~\eqref{Eq:14} and the field intensity of~\eqref{Eq:16} as the constraints to numerically calculate $P_{in,j}$ and $P_{out,j}$. From the constraints, we consider $\beta,\ I_S,\ X,\ z_{R,0}',\ z_{R,2}',\  z_{0,0}'$ and $z_{0,2}'$ as free parameters in the LSM. The parameters are embedded in~\eqref{Eq:17}, which fits the expe\-rimental Z-scan curves in Figures~\ref{Fi:6} and~\ref{Fi:7}. The values of the fitting parameters are presented in Table~\ref{Ta:4}.\\ 
We assume that the nonlinear refractive index in the sample produces a lens effect. From~\eqref{Eq:7}, we notice that $\gamma$ modifies the propagation parameters in both modes (focal plane position and Rayleigh parameter), with respect to the measured values in free-space pro\-pagation (see Table~\ref{Ta:3}). Thus, we treat them as free parameters in the fitting. We can also calculate the nonlinear refractive index for the main mode using~\eqref{Eq:8}, and the values of $z_{R,0}$ and $z_{R,0}'$ from Table~\ref{Ta:3} and Table~\ref{Ta:4} respectively. The $\gamma$ values are: $9.2\times 10^{-14}$ cm$^2$/W for Sample A, and $8.8\times 10^{-14}$ cm$^2$/W for Sample B. These values are due to thermal lensing contributions among others.   

\begin{table}[h!]
\begin{center}
\caption{Nonlinear parameters and evolved beam pro\-pagation parameters at $\lambda_p$.} \label{Ta:4}
\begin{tabular}{ccc}
\hline
	&\small   Sample A &\small  Sample B \\
\hline
\small $\beta\,(\textrm{cm/mW})$ &\small  $(2.815\pm 0.016)\times10^{-7}$ &\small  $(1.999\pm 0.022)\times10^{-8}$ \\
\small $I_s\ (\textrm{mW/cm}^2)$ &\small  $(1.65\pm 0.27)\times10^{8}$ &\small  $(3.386\pm 0.015)\times10^{6}$ \\
\small $X\ (\%)$ &\small  $7.82\pm 0.21$ &\small  $18.71\pm 0.38$ \\
\small $z_{R,0}'$ (mm) &\small  $4.786\pm 0.040$ &\small  $4.286\pm 0.044$ \\
\small $z_{R,2}'$ (mm) &\small  $3.187\pm 0.132$ &\small  $3.069\pm 0.076$ \\
\small $\Delta z_0'$ (cm) &\small  $1.815\pm 0.014$ &\small  $0.723\pm 0.009$ \\
\hline	
\end{tabular}
\end{center}	
\end{table}

\noindent The measured curves from the Z-scan experiments are shown in Fig.~\ref{Fi:6} (sample A) and Fig.~\ref{Fi:7} (sample B) together with the curves obtained from the fitting procedure previously des\-cribed; where the horizontal scale ($z$) is measured with respect to $z_{0,0}'$ value. We perform the experiments using different pump powers so that the sensitivity of the nonlinear response is increased. The values quoted in Table~\ref{Ta:4} fit the whole set of data, ${\it{i.e.}}$ including data corresponding to different pump po\-wers.\\
Our results indicate that:
\begin{itemize}
\item The parameters from the OA Z-scan technique show an indirect behavior of phenomena related to the resonant transitions at $\lambda_p$. 
\item The phenomenological description introduced to describe non-Gaussian beams properly describes the data.
\item We show evidence of mode coupling mediated by the active media, by means of the parameter $X$. We observe an improvement when the SNP are present.
\item The change in the internal dynamics due to the SNP reduces the effective intensity saturation and the third-order nonlinear absorption at $\lambda_p$.
\item The difference between the focal plane positions ($\Delta z_0'$) is modified due to thermal lensing and other effects that contribute to $\gamma$, as shown by the results reported in Tables~\ref{Ta:3} and~\ref{Ta:4}. The focal plane position for each mode evolves differently, since both modes have different intensities. 
\item Using a single $LG_{00}$ mode it is not possible to reproduce the right-side shoulder observed in Figures~\ref{Fi:6} and~\ref{Fi:7}; a proper description is achieved including the bimodal structure of the beam in the analysis. Both modes have different focal planes, while the hill (valley) is dominated by the $LG_{00}$ mode, the right-side shoulder requires the joint contribution of $LG_{00}$ and $LG_{20}$ modes. We indicate in Figures~\ref{Fi:6} and~\ref{Fi:7} the evolved focal plane positions of both modes. 
\end{itemize}
 
\section{Summary}\label{Sec:Con}
In this work we introduce a phenomenological description of the Z-scan technique, in OA configuration with a beam composed by two modes. Our description includes third-order nonlinear absorption, optical saturation and effective energy transfer between modes using the phenomenological parameter $X$. Additionally, this methodology allows us to calculate the nonlinear refractive index.  \\
We use this methodology to perform a complete cha\-racterization of the optical parameters of glass samples doped with  $Yb^{3+}-Er^{3+}$ and SNP. The analysis and interpretation of our data show evidence that the net effects of introducing the SNP under 908.6 nm pump are: the reduction of the saturation intensity, the third-order nonlinear absorption and the third-order nonlinear refractive index, and the increment of the energy transfer between laser modes.

\begin{appendix}
\section{Linear optical properties}\label{App:NA}
\begin{figure}[h!]
\centering
\includegraphics[width=8.5cm]{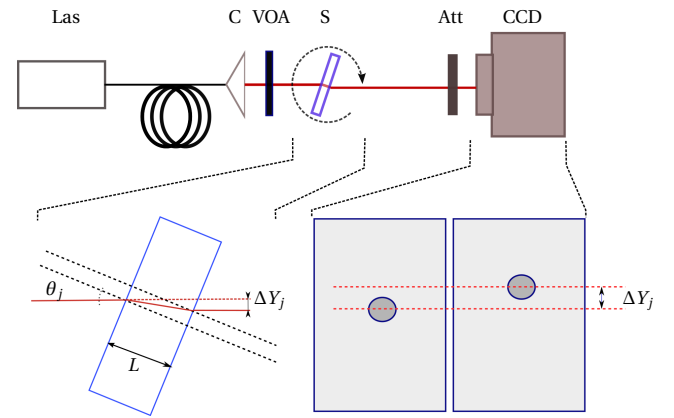} 
\caption{\textbf{Top.} Setup to measure the linear refractive index: Las - Laser; C- Asferical lens; VOA-Variable optical attenuator; S - Sample; Att- Attenuator; CCD - CCD camera. \textbf{Bottom-Left.} Sample in lateral view. \textbf{Bottom-Right.} Sketch of the profile center displacement on the CCD camera.} \label{Fi:8}
\end{figure}

In this appendix we describe the methodology to measure the linear refractive index and absorption coefficient:\\ 
\noindent \textbf{Refractive index.} The setup and its operation principle are depicted in Figure~\ref{Fi:8}. The collimated laser beam passes through the sample, which is placed at a rota\-ting sample holder. When the sample is rotated, the output beam exhibits a transverse displacement of its profile center. In order to measure this displacement, the output beam profile is recorded with a CCD ca\-mera with a resolution of 4.65\,$\mu$m$\times$4.65\,$\mu$m per pixel (Thorlabs DCU224M). \\   
The angles of incidence at the flat faces of the samples are $\theta_j \in \{0^\mathrm{o}, 8^\mathrm{o}, 12^\mathrm{o}, 16^\mathrm{o}, 20^\mathrm{o}, 24^\mathrm{o} \}$. The profile center displacements $\{\Delta Y_j\}$ are obtained with the following procedure: 1) normalize the beam profile images and use them as weighted distributions; 2) find the center of mass position ($(x_{0,j},y_{0,j})$) for each distribution at each angle $\theta_j$ (statistically related to the Poyn\-ting vector of the beam); 3) calculate the profile center displacement with $\Delta Y_j=\sqrt{(x_{0,j}-x_{0,0})^2+(y_{0,j}-y_{0,0})^2)}$, where the re\-ference position is $(x_{0,0},y_{0,0})$ at $\theta_0 = 0^\mathrm{o}$.

\noindent We have used a geometrical analysis and the refractive Snell law to derive the relation among the refractive index of the sample $n$, $\Delta Y_j$ and $\theta_j$,

\begin{equation}
n=n_{a}{\sin}(\theta_{j})\sqrt{1+\frac{L.^2 \cos ^{2}(\theta_{j})}{\left[L \sin (\theta_{j})-\Delta{Y_j}\right]^{2}}} \label{Eq:18}
\end{equation}
where $n_{a}$ is the air refractive index ($\approx 1$), $L$ is the thickness of the sample.\\
\textbf{Linear absorption.} The linear absorption coefficient is measured using the collimated beam, as in the setup of Figure~\ref{Fi:8}, using normal incidence ($\theta=0^\mathrm{o}$) between the beam and sample. A set of $P_{in,i}$ and $P_{out,i}$ is measured to get $T_1$, by means of the VOA. 

\section{Beam Profile Characterization}\label{App:BPC}
We analyze the beam propagation between lenses L1 and L2 in the Z-scan setup (see Figure~\ref{Fi:3}), where a CCD camera is placed on the $z$-displacement stage. We obtain a beam profile image shifting the CCD camera in steps of 3 mm in $+z$-direction. We consider the position of the initial image as $z=0\ \mathrm{cm}$, so other $z$-positions are measured with respect to this position. 

\noindent The Laguerre-Gauss basis is the proper basis for cylindrical waveguides, like optical fibers. LG modes conserve their angular momentum after a nonlinear inte\-raction in isotropic media~\cite{Zhang2006}. Since the pigtail of the laser diode is single mode but rather short, it cannot completely filter the profile mode. In the profile ana\-lysis we use the modal decomposition considering up to the second order in both $m$ and $l$ values,   
	
\begin{equation}
\small f(x,y,z)=\sum_{m=0}^2 \sum_{l=0}^2 a_{ml} u_{ml}^{|l|}\exp\left(-u_{ml}\right)\left[L_{m}^{|l|}\left(u_{ml}\right)\right]^2+d \label{Eq:19}
\end{equation}	      
with $u_{ml}= 2(r_{ml}/w_{ml}(z))^{2}$; $a_{ml}$ is proportional to the field intensity of the $LG_{ml}$ mode; $w_{ml}(z)$ is the beam waist of each mode; $r_{ml}^2=(x-x_{0,ml})^{2}+(y-y_{0,ml})^{2}$; $(x_{0,ml},y_{0,ml})$ is the geometric center of each mode (we consider a deviation from the concentric mode geometry in the characterization of the beam parameters); $d$ is the CCD background noise. 

\noindent After a 3D fitting of the beam profiles using~\eqref{Eq:19}, we find that the beam is composed with two Laguerre-Gauss modes with different beam propagation parameters. These two modes with non-zero coefficient $a_{ml}$ are the $LG_{00}$ and $LG_{20}$. $LG_{m0}$ modes involve the zero order associate Laguerre functions, which are identical to Laguerre functions ($L_m^{{0}}(x)=L_m(x)$). Since both modes have only $m$ index, we will omit the $l$ index. In order to estimate the propagation parameters of the modes ($z_{R,m}$ and $z_{0,m}$), we have fit these values from the propagation equation, u\-sing the set  values $\{z,\ w_{m}(z)\}$. Since the modal power, $P_{m}\propto  a_{m}w_{m}^{2}$, the power fraction of each mode is calculated by,
\begin{equation}
f_{m}=\frac{P_{m}}{P_{1}+P_{2}}=\frac{a_{m}w_{m}^{2}}{a_{1}w_{1}^{2}+a_{2}w_{2}^{2}}. \label{Eq:20}
\end{equation}

\end{appendix}

\section*{Funding Information}

Consejo Nacional de Ciencia y Tecnología (CONACyT) (\textbf{PDCPN 2015-624} and \textbf{RedTC}).

\bibliography{bibZscan}

\end{document}